\documentclass[11pt]{article}
\usepackage{a4wide}
\usepackage{amsmath, amssymb, amscd, amsthm, amsfonts}
\usepackage[utf8]{inputenc}
\usepackage{graphicx}
\usepackage{hyperref}
\usepackage{cite}
\usepackage[dvipsnames]{xcolor}

\renewcommand{\d}{{\rm d}}

\newcommand{\Tr}{{\rm Tr}}

\begin{document}
{\renewcommand{\thefootnote}{\fnsymbol{footnote}}
		
\begin{center}
{\LARGE Spontaneous symmetry breaking in the BFSS model:\\[2mm]
Analytical results using the Gaussian expansion method} 
\vspace{1.5em}

Suddhasattwa Brahma$^{1}$\footnote{e-mail address: {\tt suddhasattwa.brahma@gmail.com}}, 
Robert Brandenberger$^{2}$\footnote{e-mail address: {\tt rhb@hep.physics.mcgill.ca}}, and 
Samuel Laliberte$^{2}$\footnote{e-mail address: {\tt samuel.laliberte@mail.mcgill.ca}}
\\
\vspace{1.5em}
$^1$Higgs Centre for Theoretical Physics, School of Physics and Astronomy,\\ University of Edinburgh, Edinburgh EH9 3FD, UK\\

$^2$Department of Physics, McGill University, Montr\'{e}al, QC, H3A 2T8, Canada\\

\vspace{1.5em}
\end{center}
}
	
\setcounter{footnote}{0}

\begin{abstract}
	\noindent We apply the Gaussian expansion method to the BFSS matrix model in the high temperature limit. When the (Euclidean) BFSS action is expanded about a Gaussian ansatz, it is shown that the SO(9) symmetry is spontaneously broken, analogous to what happens in the IKKT model. The analysis of the free energy, using the set of gap equations which determines the width of the Gaussian terms, is sufficient to show that this symmetry breaking happens \textit{only} when the fermionic terms are included and is absent in the bosonic case.
\end{abstract}

\section{Introduction}

Understanding the quantum origin of space and time in the very early universe requires starting from a consistent non-perturbative theory which is complete in the ultraviolet (UV), \textit{i.e.} in the high energy limit. String theory is the most promising candidate for such a theory.  To understand string theory at the high densities of the very early universe,  it is crucial to start from a non-perturbative approach\footnote{Effective field theory approximations run into serious conceptual problems (see e.g. \cite{RHBrev} for an overview), and effective field theories consistent with string theory are constrained by several criteria such as the ``swampland'' conditions \cite{swamp} (see \cite{swamprevs} for recent reviews) and the ``Trans-Planckian Censorship Conjecture'' \cite{TCC}.}. In the late 1990s it was realized that large $N$ matrix models could provide non-perturbative definitions of string theory.  Two key examples are the BFSS \cite{BFSS} and the IKKT \cite{IKKT} matrix models.

Recently, there has been a new proposal to understand the origin of space and time in the context of these matrix models \cite{us2}.  It has been shown how, starting from abstract matrix degrees of freedom, one not only obtains emergent space and time, but one can also extract variables which meaningfully depict the dynamical history of our universe. The resulting scenario has aspects in common with \textit{String Gas Cosmology} (SGC) \cite{BV}, a model of early time cosmology in which the universe was taken to begin in a high temperature equilibrium state of a gas of strings, and the thermal fluctuations were shown \cite{us1} to lead to scale-invariant spectra of both cosmological perturbations and gravitational waves, again like what results in SGC \cite{SGflucts}. What was missing in SGC, however, was an embedding of the model into a consistent dynamical framework coming from string theory.  Our present model provides such an embedding \footnote{See also \cite{Vafa} for another approach to an emergent early universe scenario based on fundamental principles of string theory which connects with SGC.}

Let us summarize our new findings in a bit more detail. Firstly, it was shown in \cite{us1} that, assuming a thermal state of the BFSS matrix model, one can extract an early universe cosmology in which the thermal fluctuations lead to roughly scale-invariant spectra for both the primordial gravitational waves and the (observable, infrared) curvature fluctuations. Moreover, the amplitude of these perturbations is given by the ratio of the string scale to the Planck scale, exactly as in SGC, and agrees with current observations for a string scale of the order of that of Grand Unification (a string scale which in agreement with what is expected from string particle phenomenology in heterotic superstring theory \cite{GSW}). The remarkable bit about this result is that it arises solely due to the form of the BFSS Lagrangian itself and the result is independent of any fine-tuning (as compared with the freedom offered by the choice of the scalar field potential for inflation). The choice of a thermal state for primordial perturbations is also natural, as we work in the Euclidean BFSS model at high temperatures. The main leap of faith in our model was that, at high temperatures, the BFSS model was assumed to exhibit a symmetry-breaking pattern in which there would emerge exactly three large spatial dimensions. The intuition behind this assumption comes from the IKKT model, where it has been long demonstrated \cite{IKKT-PT} that a large $(3+1)$-d universe emerges from the full $10$-d theory due to a spontaneous symmetry breaking (SSB) of the isometry group (see \cite{IKKTnew} for more recent results). In fact, this was one of the motivating factors for choosing a high temperature state in the BFSS model, since it is well known \cite{BFSS-IKKT} that the BFSS model approaches the IKKT model as $T \rightarrow \infty$\footnote{Coincidentally, it is indeed in this limit that we have a well-defined perturbative expansion for the thermodynamic quantities of interest for the BFSS model since the perturbation parameter $g^2N/T^3$ is small in this limit.}. 

Although it was well-motivated, there were a couple of strong assumptions in the aforementioned calculation. Firstly, at high temperatures, the fermions decouple in the BFSS model and one only recovers the bosonic IKKT action in this limit. On the other hand, it has been argued from different directions that the presence of fermions is crucial for the $SO(10)$ symmetry-breaking in the IKKT model. Secondly, there has been a combination of analytical and numerical results which have demonstrated that although all the eigenvalues of the bosonic matrices start out small (on the string scale) for the IKKT model, the SSB leads to $3$ of them becoming large after the symmetry-breaking time. In the absence of any such evidence for the BFSS model, it is a rather strong assumption to say that there is a similar dynamical explanation happening for the BFSS action.

It is precisely this lacuna which we seek to (at least, partially) fill in this paper. We will show using well-established methods that there is a SSB of the $SO(9)$ group for the Euclidean BFSS model, in the presence of fermions. This will be done by employing the Gaussian expansion method, \cite{Gaussian_methods_BFSS}  in which we will approximate the BFSS Lagrangian with a Gaussian action and then systematically calculate the corrections to this ansatz, by expanding about the Gaussian term.  The Gaussian action contains a host of free parameters (or ``masses'') which are determined by solving the so-called `gap' or `self-consistency' equations. When truncated to any finite order, this constitutes an approximation, as mentioned above. To put it differently, in this method one adds and subtracts a Gaussian piece $S_0$ to the action and then regards $S+S_0$ as the classical action and the $-S_0$ as the one-loop counter-term \cite{Lowe}. Our goal is to then calculate the free energy for such an expanded action, and see if it gets minimized for a specific choice of the Gaussian variables. Naturally, for any finite truncation, the result of the Gaussian expansion will depend on the ``mass'' parameters in the ansatz. Nevertheless, in a lot of models which are exactly solvable, it has been shown that there exist regions of parameter space where the action is independent of the Gaussian parameters. These regions signal local minima of the effective action and identifying the region with the smallest free energy enables one to predict the true vacuum of the theory. 

In order to examine the symmetry-breaking pattern of the model, we will start with an ansatz that preserves the $U(N)$ gauge group of the BFSS model but not the isometry group. The goal is to show that there exists a region of parameter space where the free energy is minimized, and that this occurs for a configuration which corresponds to the breaking of $SO(9)$ to $SO(6)$.  In this work, we will be content with studying the gap equations which deal with the bosonic Gaussian parameters\footnote{By this term, we will throughout refer to the ``mass'' terms assigned to the bosonic matrices of the model since the spatial directions emerge from the eigenvalue distribution of these matrices in the BFSS model.}, and show that such a gap equation cannot be solved by a $SO(9)$-symmetric solution in the presence of fermionic terms. In other words, we will demonstrate that as long as there are some fermionic contributions to the Gaussian ansatz, there must be a SSB of the BFSS model. Our result mimics previous computations of the IKKT model although the presence of an intrinsic time parameter complicates the application of the Gaussian method to the BFSS action. Thus, we find first evidence of SSB in the BFSS model which is at the same footing as in the IKKT model.

In the next section, we review the results of the applying the Gaussian expansion method to the IKKT model in order to introduce this technique and highlight the minimum ingredients required for showing that there must be a SSB in such a matrix model. In Sec-3, we show that, analogous to what happens in the IKKT model, there is no symmetry breaking for the bosonic BFSS action. Finally, we present our main result in Sec-4 demonstrating why there \textit{must} be SSB for the full BFSS model in the presence of fermions. We conclude in Sec-5, highlighting future directions and why our result has deep consequences for matrix cosmology.

\section{Review of the IKKT results}

Let us briefly review the results of applying the Gaussian expansion technique to the Euclidean IKKT model \cite{IKKT-GE, IKKT-not}.  The main goal of this section is to show what the precise requirements are to obtain spontaneous breaking of the $SO(10)$ symmetry.  For the IKKT model, the analytical arguments have been supplemented with numerical studies \cite{IKKT-PT} which show that the $SO(10)$ symmetry in fact breaks to some $SO(4)$ group (cross some other symmetry group for the internal dimensions). We are at the moment unable  to carry out the full analyses for the BFSS model,  since it would require a considerable amount of numerics to do so. As a result, we are not yet in the position to determine the end state of the symmetry breaking.  Nevertheless, by identifying the necessary ingredients required to identify the symmetry-breaking, we will be able to show definitively that the $SO(10)$ symmetry must necessarily be broken in the BFSS model. 

The partition function of the Euclidean IKKT model is given by (we will closely follow the notation of \cite{IKKT-not} for consistency)
\begin{eqnarray}\label{IKKT1}
	Z = \int \d X \d \Phi \, e^{-S_{\rm IKKT}}\,,
\end{eqnarray}
where the IKKT action is defined as:
\begin{eqnarray}\label{IKKT2}
	S_{\rm IKKT} = -\dfrac{1}{4} N\, \Tr \left[X_\mu, X_\nu\right]^2 - \dfrac{i}{2} N \,\Tr \left(\Phi_\alpha \left(\tilde{\Gamma}_\mu\right)_{\alpha\beta} \left[X_\mu, \Phi_\beta\right]\right) =: S_{\rm IKKT}^{(b)} + S_{\rm IKKT}^{(f)}\,.
\end{eqnarray}
In the above expressions we have rescaled the $10$ bosonic matrices $A_\mu \rightarrow \lambda^{1/4}\, X_\mu$ and their $16$ fermionic superpartners $\Psi_\alpha \rightarrow \lambda^{3/8} \,\Phi_\alpha $ by some powers of the 't Hooft coupling $\lambda = g^2 N$, so as to make the action independent of the Yang-Mills coupling $g$. The $10-d$ gamma matrices in the Weyl basis, $\Gamma_{\mu}$, are multiplied by the charge conjugation matrix $\mathcal{C}$, to get $\tilde{\Gamma}_\mu = \mathcal{C} \Gamma_\mu$. From now on, we will assume the number of bosnic matrices to be $d$ (\textit{i.e.}, $\mu = 1,2,\ldots, d$), while the fermionic ones to be $p$ (\textit{i.e.}, $\alpha = 1,2,\ldots, p$), to be completely general. The reason for separating out the bosonic and fermionic parts of the IKKT action will be clear later on.

The general idea of the Gaussian expansion method \cite{Gaussian_methods_BFSS, Lowe} is to approximate the above action by a Gaussian ansatz which is the most general $SU(N)$-invariant action and yet does not assume an $SO(d)$ symmetry.  One then expands the action around this Gaussian ansatz and calculates quantities such as the free energy to whichever order in perturbation theory is desired,  and minimizes the free energy by solving a set of gap equations which fixes the parameters of the Gaussian terms. In general, for large-$N$ theories, it is known that planar diagrams are the only ones which contribute to the gap equations when calculating corrections to the Gaussian approximation. 

More concretely, we introduce a Gaussian action of the form
\begin{eqnarray}\label{IKKT_S0}
	S_0 = \sum^{d}_{\mu=1} \left(\dfrac{N}{v_\mu}\right) \Tr\left(X_\mu X_\mu\right) + \left(\dfrac{N}{2}\right) \sum_{a=1}^{N^2-1} \Phi_\alpha^a \, \mathcal{A}_{\alpha\beta}\, \Phi_\beta^b =: S_0^{(b)} + S_0^{(f)}\,,
\end{eqnarray}
where the Gaussian parameters are $v_\mu > 0$ can, a priori, take $d$ distinct values and $\mathcal{A}$ is a $p\times p$ matrix, and we have separated out the bosonic and fermionic Gaussian terms. \footnote{We have expanded the fermionic fields $\Phi_{\alpha}$ in terms of the $N^2 - 1$ $SU(N)$ generators, the $\Phi_{\alpha}^a$ being the expansion coefficients.} 

\subsection{No symmetry breaking for bosonic IKKT}

In order to understand the full power of the Gaussian expansion method in exploring the symmetry breaking pattern of the model, let us first apply it to the bosonic part of the IKKT model, \textit{i.e.} to the $S_{\rm IKKT}^{(b)} = -\dfrac{1}{4} N\, \Tr \left[X_\mu, X_\nu\right]^2$ part of the action. Naturally, we will now only consider the part of the Gaussian ansatz $S_0^{(b)}$ that depends on the ``bosonic mass'' parameters $v_\mu$. 

First, let us express the partition function in terms of an expansion around $S_0^{(b)}$, namely
\begin{eqnarray}
	Z^{(b)}_{\rm IKKT} = Z_0^{(b)} \, \left\langle e^{-\left(S_{\rm IKKT}^{(b)}-S_0^{(b)}\right)}\, \right\rangle_0\,,
\end{eqnarray}
where $Z_0^{(b)} \sim \int \d X e^{S_0^{(b)}}$ and the expectation values (denoted by the angled brackets) are taken with respect to $Z_0^{(b)}$ and we have ignored factors of $\lambda$ in the prefactor as they are of no consequence to us. Although one can systematically calculate the free energy of the bosonic part of the IKKT action from the partition function above $F_{\rm IKKT}^{(b)} = -\ln Z^{(b)}_{\rm IKKT}$ by expanding it in a power series \cite{IKKT-GE,IKKT-not}, for our purposes, it will be sufficient to consider only the first order correction. 

More concretely, once the bosonic matrices are expanded with respect to $SU(N)$ generators and the measure of the integrals written appropriately in terms of them, it is easy to carry out a series of arduous but straight-forward Gaussian integrals to express the free energy as\footnote{See \cite{IKKT-not} for details of these calculations. More explicitly, we need to do a bunch of Gaussian integrals after writing out the measure as $\d X = \prod_{a=1}^{N^2-1} \prod_{\mu=1}^{d} \d x^a_\mu$. Specifically, the prefactor of $(N^2-1)$ arises since the gauge group is $SU(N)$ and has $N^2 - 1$ generators. Note that we have ignored some factors of $\pi$ throughout.}:
\begin{eqnarray}\label{IKKT_bosonic_free}
	F_0^{(b)} &:=& - \ln Z_0^{(b)} = \frac{1}{2}\left(N^2 -1\right) \left[C_1  - \sum_{\mu=1}^{d} \ln v_\mu\right]\,,\\
	F_1^{(b)} &:=& \left\langle S^{(b)}_{\rm IKKT} \right\rangle_0 - \left\langle S^{(b)}_0 \right\rangle_0 = \frac{1}{8} \left(N^2 -1\right)\left[\sum_{\mu\neq\nu}   v_\mu v_\nu - C_2 \right]\,,
\end{eqnarray}
where $C_1$ and $C_2$ are constants which depend on the $\lambda$ and $d$ and contain some numerical factors.   

Our simple goal is to minimize the (bosonic) free energy, to this order, by varying the Gaussian parameters $v_\mu$, \textit{i.e.}
\begin{eqnarray}
	\dfrac{\partial}{\partial v_\mu} \left( F_0^{(b)} + F_1^{(b)}\right) = 0\,,
\end{eqnarray}
which gives us the bosonic gap equations:
\begin{eqnarray}\label{Gap_IKKT_b}
	- \frac{1}{2 v_\mu} + \frac{1}{4}\sum_{\mu\neq\nu} v_\mu = 0 \,.
\end{eqnarray}
Note that these are actually $d$ equations written in a compact form. What is important is that, given that $v_\mu \ge 0$ in our ansatz (which is still the most general Gaussian action one can choose), we find that the solution of the above equation is given by a $SO(d)$ symmetric solution:
\begin{eqnarray}
	v_1 = v_2 = \cdots = v_d = \sqrt{\dfrac{2}{d-1}}\,.
\end{eqnarray}
Therefore, although we did not start with an $SO(d)$ symmetric ansatz for the Gaussian term, the bosonic part of the IKKT action is such that the solution which minimizes the free energy is $SO(d)$ symmetric thereby indicating that the symmetry is unbroken. However, looking ahead, note that we will always look at the bosonic gap equations to tell us if there is a symmetry breaking in the theory as it is indeed the eigenvalue distribution of the bosonic matrices which is conjectured to give us the emergent spacetime in this theory.

\subsection{Symmetry breaking and the role of the fermionic terms}

Let us now go back to the full IKKT action, including the fermionic terms, and therefore include the fermionic part of the Gaussian ansatz $S_0^{(f)}$. The Gaussian parameters are encapsulated in the $p\times p$ anti-symmetric matrix $\mathcal{A}$:
\begin{eqnarray}\label{form_A}
	\mathcal{A}_{\alpha\beta} := \dfrac{i}{3!} \sum_{\mu\nu\lambda} \omega_{\mu\nu\lambda} \left(\mathcal{B}_{\mu\nu\lambda}\right)_{\alpha\beta}\,,
\end{eqnarray}
where $\mathcal{B}_{\mu\nu\lambda} \sim \mathcal{C} \Gamma_\mu  \Gamma_\nu^\dagger  \Gamma_\lambda$ and we have suppressed the spinor indices in the last expression to avoid clutter. Note that $\mathcal{A}$ could have had a term of the form $\omega_\mu \Gamma_\mu$ which is absent due to the Majorana nature of the fermions in $10$-d. However, this is not crucial for our arguments below. In fact, the explicit form of $\mathcal{A}$ will not enter our expressions and all we willneed is its $SO(d)$ index structure. 

Before going through the calculation, note that we will have to collect the expanded terms in the free energy in a slightly different way due to the fermionic terms. For instance, we have to count terms of the form $S_0^{(f)}$ and $\left(S_{\rm IKKT}^{(f)}\right)^2$ to be of the same order in the reorganized expansion \cite{GE_fermion}. Keeping this in mind, and once again expanding only to first order, one finds that
\begin{eqnarray}
	F_0 &:=& - \ln Z_0 = \frac{1}{2}\left(N^2 -1\right) \left[C_3  - \sum_{\mu=1}^{d} \ln v_\mu - \ln \left({\rm Pf} \mathcal{A}\right)\right]\,,\\
	F_1^{(b)} &:=& \left\langle S^{(b)}_{\rm IKKT} \right\rangle_0 - \left\langle S^{(b)}_0 \right\rangle_0 - \frac{1}{2} \left\langle \left(S^{(f)}_{\rm IKKT}\right)^2 \right\rangle_0\\
	&=& \frac{1}{8} \left(N^2 -1\right)\left[\sum_{\mu\neq\nu}   v_\mu v_\nu + C_4 - 4 \sum_\mu \rho_\mu v_\mu \right]\,,\nonumber
\end{eqnarray}
where the constants $C_3$ and $C_4$ have new contributions (compared to the corresponding constants $C_1$ and $C_2$ in \eqref{IKKT_bosonic_free}) coming from the fermionic terms which, like before, depend on $N, d$ and now on $p$. All the expectation values are now taken with respect to the full Gaussian action (but we still keep the same subscript $0$ to denote this). We have also introduced the Pfaffian of the Gaussian matrix $\mathcal{A}$, defined as ${\rm Pf} \mathcal{A} := {\rm det} \mathcal{A}^{1/2}$.  More importantly, we have defined
\begin{eqnarray}\label{rho}
	\rho_\mu := \frac{1}{4} \Tr \left[\left(\mathcal{A}^{-1}\Gamma_\mu\right)^2\right]\,,
\end{eqnarray}
where we have taken the trace over $p$-dimensional spinor indices $\alpha, \beta$. The details of the above terms will not be important for our argument.  We will only need to focus on the bosonic gap equation (the self-consistency equation according to the definition of \cite{IKKT-GE,IKKT-not}) at first order:
\begin{equation}
	\dfrac{\partial}{\partial v_\mu} \left( F_0 + F_1\right) = 0\,
	\end{equation} 
which yields
\begin{equation}\label{IKKT_Gap}
- \frac{1}{2 v_\mu} + \frac{1}{4}\sum_{\mu\neq\nu} v_\mu  - \frac{1}{4}\rho_\mu = 0 \,.
\end{equation}
Just by inspecting this equation \eqref{IKKT_Gap}, we can conclude that there must be a breaking of the $SO(d)$ symmetry. The argument simply depends on realizing that the $\rho_\mu$, as defined in \eqref{rho}, are different for different choices of $\mu$ due to the appearance of the $\Gamma$ function. Therefore, this bosonic gap equation cannot have a solution of the form $v_1 = v_2 = \cdots = v_d$, as before. As promised, the explicit form of the matrix $\mathcal{A}$ did not play a role in this derivation.  This point was already pointed out in \cite{IKKT-not}. 

The main reason for the symmetry breaking is the form of the interaction between the fermionic and bosonic matrices, as encoded in $S_{\rm IKKT}^{(f)}$ in \eqref{IKKT2}. The appearance of the $\Gamma$ function in that term is the reason why the bosonic gap equation gets a ``source'' term for the $v_\mu$ which is not $\mu$-independent What is important for us is that inspecting the bosonic gap equation is sufficient to detect the presence (or absence) of symmetry-breaking. Of course, the contributions of the fermionic terms to this gap equation is what turned out to be crucial. In fact, symmetry breaking is guaranteed as long as $\mathcal{A}$ has some non-zero entries, \textit{i.e.} the Gaussian parameters $\omega_{\mu\nu\lambda}$ are not all zero. And that is why we never even needed to write down the fermionic gap equation to draw this conclusion. 

Of course, without inspecting the full system of gap equations, it will not be possible to conclude what the $SO(d)$ symmetry break into. In the case of the IKKT model,  numerical tools were required to demonstrate what the symmetry of the solution which minimizes the free energy is.  But at the moment we do not have the required tools to study this question for the BFSS model, at least not in this work.  It would be nice to gain a good physical understanding for why the resulting symmetry of the Lorentzian matrix model is $SO(3)$, and why the solution corresponds to three expanding dimensions while six remain microscopical. This is a question we are currently working on. But in the following, we will address the restricted question of showing, in analogy with what was described above in the case of the IKKT model, that there is symmetry breaking in the BFSS model as long as the contributions of the fermionic terms are taken into account.

\section{Gaussian expansion method for the bosonic BFSS model}

Just like for the Euclidean IKKT model, we will start with the bosonic BFSS action first and apply the Gaussian expansion technique to it.  As we described in the case of the symmetry breaking analysis in the IKKT model, we will not assume a $SO(D)$ symmetric ansatz to begin with and will consider the most general $U(N)$-invariant Gaussian ansatz\footnote{There is going to be a minor difference between the calculations done here compared to what was done in the previous section in the choice of the gauge group. While we chose it to be $SU(N)$ in the case of the IKKT analysis summarized above, we will stick to $U(N)$ for the BFSS model as was done in \cite{Gaussian_methods_BFSS}.} which is, however, not $SO(D)$ symmetric. 

The bosonic part of the Euclidean BFSS action is given by (now we follow the notation of \cite{Gaussian_methods_BFSS} for easy comparison):
\begin{eqnarray}\label{BFSS_bosonic1}
	S_{\rm BFSS}^{(b)} = \frac{1}{g^2} \int\d\tau \, \Tr\left\{\frac{1}{2} D_\tau X^i D_\tau X^i - \frac{1}{4} \left[X^i, X^i\right]^2\right\}\,.
\end{eqnarray}
There are $D = 9$ $SU(N)$ bosonic matrices for the BFSS model, denoted by $X^i$ above. We have changed the indices from Latin $\mu$ to Roman $i$ to indicate that $i = 1, 2, \ldots, d-1$, in the notation of the previous section. Let us keep the dimension arbitrary as before and denote $D =d-1$ to make connection with the previous section. $\tau$ here denotes the Euclidean time direction and $\beta := 1/\tau$. We would typically be interested in the high-temperature (confined) behaviour of the model since this is the regime which is interesting for early-universe cosmology \cite{us1}. 

The first new complication of the BFSS model is to have a gauge-covariant derivative $D_\tau := \partial_\tau + i \left[A_0, \cdot\right]$. We fix the gauge as $\partial_\tau A_0 = 0 \, \Rightarrow \, A_0 := A_{00}/\sqrt{\beta} = {\rm const.}$.  On introducing the necessary ghost fields $\alpha, \bar{\alpha}$, we can write the action as:
\begin{eqnarray}
	S_{\rm BFSS}^{(b)} = \frac{1}{g^2} \int\d\tau \, \Tr\left\{\frac{1}{2} \partial_\tau X^i \partial_\tau X^i - \frac{1}{2} \left[A_0, X^i\right]^2 + i \left[A_0, X^i\right] \left(\partial_\tau X^i\right) - \frac{1}{4} \left[X^i, X^i\right]^2 + \partial_\tau \bar{\alpha} D_\tau\alpha  \right\}\,.
\end{eqnarray}
Then we can Fourier expand all the fields (in their Matsubara frequencies in units of $\omega = 2\pi/\beta$) as 
\begin{eqnarray}
	X_i(\tau) = \sum_l X_l^i \, e^{i l \omega \tau}\,,\;\;\;\;\; \alpha(\tau) = \sum_{l\neq 0} \alpha_l e^{i l \omega \tau}\;\;\; {\rm and}\;\;\;\;\; \bar{\alpha}(\tau) = \sum_{l\neq 0} \bar{\alpha}_l e^{-i l \omega \tau}\,,
\end{eqnarray}
using which, we can expand the BFSS action to get
\begin{eqnarray}\label{BFSS_bosonic}
	S_{\rm BFSS}^{(b)} &=& \frac{1}{2 g^2} \sum_l \left(\dfrac{2\pi l}{\beta}\right)^2 \Tr\left(X_l^i X_{-l}^i\right) + \frac{1}{g^2} \sum_{l\neq 0} \left(\dfrac{2\pi l}{\beta}\right) \Tr\left(\bar{\alpha}_l \alpha_{l}\right)\\
	& & - \frac{1}{g^2 \sqrt{\beta}}\sum_l \left(\dfrac{2\pi l}{\beta}\right)^2 \Tr\left(X_l^i \left[A_{00}, X_{-l}^i\right]\right) + \frac{1}{g^2 \sqrt{\beta}}\sum_{l\neq 0} \left(\dfrac{2\pi l}{\beta}\right) \Tr\left(\bar{\alpha}_l \left[A_{00}, \alpha_{-l}\right]\right) \nonumber\\
	& & +  \frac{1}{2 g^2 \beta}\sum_l \Tr\left(\left[A_{00}, X_l^i\right] \left[A_{00}, X_{-l}^i\right]\right) -   \frac{1}{4 g^2 \beta}\sum_{l+m+n+p=0} \Tr\left(\left[X_l^i, X_m^j\right] \left[X_n^i, X_p^j\right]\right)\nonumber\,.	
\end{eqnarray}
It is easy to see that the bosonic part of the Euclidean BFSS action, written as above, is the same as the one used in \cite{us1,  BFSS-IKKT} when we rescale some the fields by some factors of $\beta$. Moreover, the sums over the Fourier modes are arranged in a slightly different way than in \cite{BFSS-IKKT} , and the Mastsubara frequencies have been written out explicitly. Finally, the 't Hooft coupling has not been set to one, as is typically done. 

Let us now introduce our Gaussian ansatz for the above bosonic action as:
\begin{eqnarray}\label{BFSS_Gaussian_bosonic}
	S_0^{(b)} = - \frac{N}{\Lambda} \Tr \left(U + U^\dagger\right) + \sum_l \sum_{i=1}^{D} \dfrac{1}{2 v_{l,i}} \Tr \left(X_l^i X_{-l}^i\right) - \sum_{l\neq 0}\frac{1}{s_l} \Tr\left(\bar{\alpha}_l\alpha_l\right)\,,
\end{eqnarray}
where $U:= \mathcal{P} e^{i \oint \d\tau A_0} = e^i\sqrt{\beta} A_{00}$ is the holonomy of the gauge connection $A_{00}$ and takes its value in the $SU(N)$ group. Clearly, the first term corresponding to $U$ is not a quadratic term, and so the action is not strictly-speaking a Gaussian one. However, this is the appropriate term to include for angular variables as argued in \cite{Gaussian_methods_BFSS, Lowe}. The Gaussian parameter corresponding to this is given by $\Lambda$ whereas we keep calling the `bosonic mass' parameters $v_{l,i}$ in accordance with the previous section (and differing from what has been done in \cite{Gaussian_methods_BFSS}).. However, recall that $v_{l,i}>0$ for us. More importantly, we have added an $i$ index to this Gaussian parameter to allow for a breaking of the $SO(D)$ symmetry, as was done for the IKKT model. This is a necessary new generalization for us (say, as compared to \cite{Lowe}) since we are interested in studying symmetry-breaking. However, as we will show later, we can derive the $SO(D)$-symmetric solution of \cite{Gaussian_methods_BFSS}, which was found by studying the gap equations in the bosonic case. 

From a practical point of view, what complicates the case for the BFSS model is the appearance of the $l$ index due to the presence of an intrinsic time parameter which necessitates expansion in terms of Fourier modes. Finally, the ghost fields have the usual form, with $s_l$ denoting the Gaussian parameters. Already before going into the calculations, we can see that the gap equations even for the purely bosonic case will be of three types -- one for the Wilson loops, one for the bosonic matrices and one for the ghost terms. Contrast this with the single type of gap equations we had for the bosonic IKKT model \eqref{Gap_IKKT_b}. However, as we will show explicitly below, some of the gap equations will decouple and allow us to study the bosonic gap equations, relevant for discovering symmetry-breaking patterns, in isolation. 

The necessary ingredients for doing the calculations are to calculate propagators for the bosonic and ghost field, as well as the expectation values of the Wilson-loop operators. Let us begin by diagonalizing the holonomy $U = {\rm diag.} \left(e^{i \alpha_1}, e^{i \alpha_2}, \ldots, e^{i \alpha_N}\right)$, and then quoting the one-plaquette partition function for the holonomy \cite{GW, Lowe}:
\begin{eqnarray}\label{Hol_1}
	Z_{\square} = e^{-\beta F_\square} = 
	\begin{cases}
		\exp{\left(N^2 \left(-\frac{2}{\Lambda} - \frac{1}{2} \ln \frac{\Lambda}{2} +\frac{3}{4}\right)\right)}& \;\;\;\;\;\;\;\; \Lambda \leq 2\\
		\exp{\left(-N^2/\Lambda^2\right)}& \;\;\;\;\;\;\;\; \Lambda \ge 2\,,
	\end{cases}
\end{eqnarray}
with a phase-transition at $\Lambda =2$. Similarly, the expectation value of the Wilsonian loop is 
\begin{eqnarray}\label{Hol_2}
	\left\langle \Tr\, U \right\rangle_\square =
	\begin{cases}
		N(1 -\Lambda/4) & \;\;\;\;\;\;\;\; \Lambda\leq 2\\
		N/\Lambda & \;\;\;\;\;\;\;\; \Lambda \ge 2\,.
	\end{cases}
\end{eqnarray}
These will be useful for calculating the free energy to the first order. Next, it is easy ot calculate the propagators of the gauge field, the bosonic and the ghost fields (using the Gaussian action \eqref{BFSS_Gaussian_bosonic}):
\begin{eqnarray}\label{Gaussian_prop_bosonic}
 \left\langle \left(A_{00}\right)_{AB} \left(A_{00}\right)_{CD}\right\rangle_0 &=& \rho_0^2\, \delta_{AD} \,\delta_{BC}\,,\nonumber\\
 \left\langle \left(X_l^i\right)_{AB} \left(X_m^j\right)_{CD}\right\rangle_0 &=& v_{l,i}\, \delta^{ij}\, \delta_{l+m}\, \delta_{AD}\, \delta_{BC}\,,\nonumber\\
 \left\langle \left(\bar{\alpha}_l\right)_{AB} \left(\alpha_m\right)_{CD}\right\rangle_0 &=& s_l\, \delta_{lm}\, \delta_{AD}\, \delta_{BC}\,,
\end{eqnarray}
where the expectation values are taken with respect to the Gaussian action in \eqref{BFSS_Gaussian_bosonic}, and we have defined $\rho_0$ in terms of the eigenvalues of the holonomy $\alpha$ as:
\begin{eqnarray}
	\rho_0^2 = \frac{1}{\beta N} \int \d \alpha \, \alpha^2 \rho_\square(\alpha)\,.
\end{eqnarray}
The explicit expression for the above can be found in \cite{Gaussian_methods_BFSS} but this is not important for our purposes. 

Before going ahead with the calculation, let us make one observation. In the IKKT model \eqref{IKKT2}, there are no quadratic (or kinetic) terms. However, this is not the case in the BFSS model as can be seen from the presence of terms of the form:
\begin{eqnarray}
\frac{1}{g^2} \left[\frac{1}{2} \sum_l \left(\dfrac{2\pi l}{\beta}\right)^2\, X_{-l}^i X_l^i + \sum_{l\neq 0} \left(\dfrac{2\pi l}{\beta}\right)^2 \, \bar{\alpha}_l \alpha_l \right]\,.
\end{eqnarray}
Naturally, these terms will also give contributions to the propagators for the bosonic and the ghost field, of the form:
\begin{eqnarray}\label{BFSS_bosonic_prop}
	\left\langle \left(X_l^i\right)_{AB} \left(X_m^j\right)_{CD}\right\rangle &=& \left(\dfrac{g \beta}{2\pi l}\right)^2\, \delta^{ij}\, \delta_{l+m}\, \delta_{AD}\, \delta_{BC}\,,\nonumber\\
	\left\langle \left(\bar{\alpha}_l\right)_{AB} \left(\alpha_m\right)_{CD}\right\rangle &=& \left(\dfrac{g \beta}{2\pi l}\right)^2\, \delta_{lm}\, \delta_{AD}\, \delta_{BC}\,,
\end{eqnarray}
where the form of the delta functions $\delta_{AD}\delta_{BC}$ can be understood by writing out the matrices explicitly in some basis of generators of the $U(N)$ group. The angled brackets above refer to calculating the expectation values with respect to the BFSS partition function, and not with respect to the Gaussian one. Crucially for us, to the next-to-leading order that we will consider for calculating the free energy, we will not need the above propagators to do the computation. More explicitly, for evaluating $\left(F_0 + F_1\right)$, we only need $\left\langle S\right\rangle_0$ and $\left\langle S_0 \right\rangle_0$ -- none of which will require going beyond the propagators given in \eqref{Gaussian_prop_bosonic}.

Given this background, we can write the free energy to zero'th order as:
\begin{eqnarray}\label{F0_bosonic}
	\beta F_0 = \beta F_\square(\Lambda) - \frac{N^2}{2}\sum_{i=1}^{D} \sum_l \ln v_{l,i} + N^2 \sum_{l\neq 0} \ln s_l\,.
\end{eqnarray}
The details of this calculation has been delegated to Appendix \eqref{App1}. To calculate the first order correction to the free energy, we recall that
\begin{eqnarray}\label{F1_bosonic1}
	F_1 = \left\langle\left(S - S_0\right)\right\rangle_0\,.
\end{eqnarray}
Once again, we have left the details to  Appendix \eqref{App2}, and only quote the final result here:
\begin{eqnarray}\label{F1_bosonic}
F_1 &=& F_\square^{(1)} \left(\Lambda\right) + \frac{N^2}{2} \sum_{i=1}^{D} \sum_l \left[\frac{1}{g^2} \left(\frac{2\pi l}{\beta}\right)^2 v_{l,i}\right] - \frac{N^2 D}{2} \sum_l \, 1 \nonumber \\ 
& & + N^2 \sum_{l\neq 0} \left[\frac{1}{g^2} \left(\dfrac{2\pi l}{\beta}\right)^2 \, s_l + 1\right] + \dfrac{N^3}{g^2 \beta}\, \rho_0^2 \,\sum_{i=1}^D \sum_l v_{l,i} +  \dfrac{N^3}{2 g^2 \beta} \underbrace{\sum_{i=1}^D \sum_{j=1}^D}_{j\neq i}\, \sum_{l,k} v_{l,i}\,v_{k,j}\,,
\end{eqnarray}
where 
\begin{eqnarray}\label{F1_hol}
	F_\square^{(1)}  \left(\Lambda\right) = \frac{N}{\Lambda} \left\langle\Tr\, (U +U^\dagger)\right\rangle_\square =
	\begin{cases}
		\dfrac{2 N^2}{\Lambda} \left(1-\frac{\Lambda}{4}\right)\,,&\;\;\;\; \Lambda\leq 2\\[2mm] 
		\dfrac{N^2}{\Lambda^2}\,,&\;\;\;\; \Lambda \geq 2
	\end{cases}
\end{eqnarray}

Once we have the free energy calculated to the next-to-leading order, we can evaluate the gap equations as:
\begin{eqnarray}\label{BFSS_gap_bosonic}
	\frac{\partial}{\partial v_{l,i}} \left(F_0 + F_1\right) &=& 0\\
	\frac{\partial}{\partial s_l} \left(F_0 + F_1\right) &=& 0\\
	\frac{\partial}{\partial \Lambda} \left(F_0 + F_1\right) &=& 0\,.
\end{eqnarray}
While the first equation is the one of interest for us, as it gives the gap equation for the bosonic Gaussian parameter, the second one refers to the gap equation for the ghosts and the third one for the Wilson loop parameters. Since the full free energy takes the form
\begin{eqnarray}\label{F1}
	F_0 + F_1 &=& \beta F_\square(\Lambda) - \frac{N^2}{2}\sum_{i=1}^{D} \sum_l \ln v_{l,i} + N^2 \sum_{l\neq 0} \ln s_l + \frac{N}{\Lambda} F_\square^{(1)} \left(\Lambda\right)   + \frac{N^2}{2} \sum_{i=1}^{D} \sum_l \left[\frac{1}{g^2} \left(\frac{2\pi l}{\beta}\right)^2 v_{l,i}\right] \\
	 & & - \frac{N^2 D}{2} \sum_l \, 1 + N^2 \sum_{l\neq 0} \left[\frac{1}{g^2} \left(\dfrac{2\pi l}{\beta}\right)^2 \, s_l + 1\right] + \dfrac{N^3}{g^2 \beta}\, \rho_0^2 \,\sum_{i=1}^D \sum_l v_{l,i} +  \dfrac{N^3}{2 g^2 \beta} \underbrace{\sum_{i=1}^D \sum_{j=1}^D}_{j\neq i}\, \sum_{l,k} v_{l,i}\,v_{k,j} \,,\nonumber
\end{eqnarray}
the gap equations for the ghosts can be written as:
\begin{eqnarray}\label{BFSS_gap_ghost}
	\frac{N^2}{s_l} + N^2 \left(\dfrac{2 \pi l}{\beta}\right)^2\, \frac{1}{g^2} = 0\;\;\;\;\;
\Rightarrow	\;\;\; \dfrac{g^2}{s_l} = - \left(\dfrac{2 \pi l}{\beta}\right)^2\, \;\; \;\;\; l\neq 0\,.
\end{eqnarray}
This shows that, to this order, the ghost fields are completely decoupled from the rest as their solutions are free from the other parameters. This, of course, greatly simplifies our analysis. The gap equations obtained by varying  the holonomy parameter $\Lambda$ can be written as\footnote{The reason why the extent of space parameter appears in the gap equation for $\Lambda$ is that there is a term in \eqref{F1} which involves both $\rho_o(\Lambda)$ and $v_{l,i}$.}:
\begin{eqnarray}\label{BFSS_gap_hol}
	-\left(1 - \frac{2}{\Lambda}\right) \ln \left(\dfrac{1-\Lambda}{2}\right) + 1 &=& \dfrac{N g^2 \beta}{4 \sum_{i=1}^D \left\langle (R^i)^2\right\rangle}\,,\;\;\;\;\;\; \Lambda\leq 2\\
	\Lambda &=& \dfrac{N g^2 \beta}{2 \sum_{i=1}^D \left\langle (R^i)^2\right\rangle}\,,\;\;\;\;\;\; \Lambda\geq 2\,,
\end{eqnarray}
where we have introduced the `extent of space parameters' for the eigenvalue distribution of the bosonic matrices, and therefore gives an estimate of the ``size'' of the system, as:
\begin{eqnarray}
\left\langle (R^i)^2\right\rangle := \frac{1}{N} \Tr \, \left\langle \left(X^i(\tau)\right)^2\right\rangle = \frac{N}{\beta} \sum_l\, v_{l,i}\,.
\end{eqnarray}
Importantly, note that there is no sum over the $i$ index in the above equation. These are the same extent of parameters which one uses in extracting cosmological solutions from the BFSS model and when spatial isotropy of the large dimensions is assumed (which is why we will simply get factors of $D$ in the denominator in front of $\langle (R^i)^2\rangle$ in \eqref{BFSS_gap_hol}). We will have more to say about this later on. However, for now note that the bosonic parameters $v_{l,i}$ are indeed coupled to $\Lambda$ through the term $N^3/(g^2 \beta)\, \rho_0^2 \,\sum_{i=1}^D \sum_l v_{l,i}$ in the free energy.

Finally, let us derive the gap equations for $v_{l,i}$, which are given as
\begin{eqnarray}\label{BFSS_bos_gap1}
	- \frac{1}{v_{l,i}} + \frac{1}{g^2} \left(\dfrac{2\pi l}{\beta}\right)^2 + \dfrac{2 N}{g^2 \beta}\, \rho_0^2 (\Lambda)  + \dfrac{2 N}{g^2 \beta} \,\sum_{\j\neq i} \sum_l \, v_{l,j} = 0\,.
\end{eqnarray}
Although we did not choose a Gaussian ansatz which is $SO(D)$ symmetric to begin with, we can now search for solutions of the above equations to examine if $SO(D)$ symmetric solutions are allowed or not. Notice the remarkable similarity between this equation and the gap equation for the bosonic IKKT action given in \eqref{Gap_IKKT_b}. This encourages us to look for solutions of the form $v_{l,i} = v_{l,2} = \cdots = v_{l,D} =: v_l \,, \; \forall\; l \in \mathbb{Z}$, and plugging this ansatz into \eqref{BFSS_bos_gap1}, we find
\begin{eqnarray}\label{BFSS_bos_gap}
	\frac{g^2}{v_{l}} =  \left(\dfrac{2\pi l}{\beta}\right)^2 + \dfrac{2 N}{\beta}\, \rho_0^2 (\Lambda)  + \dfrac{2 N (D-1)}{\beta} \, \sum_l \, v_{l}\,.
\end{eqnarray}
This is the same gap equation which had been derived in \cite{Gaussian_methods_BFSS}  assuming an $SO(D)$-symmetric Gaussian action. Therefore, we have proved that the $SO(D)$ group remains unbroken for the bosonic BFSS action\footnote{Strictly speaking, we have proved that our more general gap equation allows for $SO(D)$-symmetric solutions and we have not proven the uniqueness of the ensuing solution. However, given that the gap equation is an algebraic one, it easily follows that the $SO(D)$-symmetric solution of \cite{Gaussian_methods_BFSS} is the only one.}. Note that our proof neither requires the explicit form of $\rho_0 (\Lambda)$ nor the analyses of the other gap equations \eqref{BFSS_gap_ghost} and \eqref{BFSS_gap_hol}, as was hinted earlier on.

However, for the sake of completeness, we will go on to show the solutions of \eqref{BFSS_bos_gap}, which can be written as (following the notations of \cite{Gaussian_methods_BFSS}):
\begin{eqnarray}
	v_l = \dfrac{g^2}{\left(\frac{2\pi l}{\beta}\right)^2 + m_{\rm eff}^2}\,,
\end{eqnarray}
where the effective thermal mass is defined as
\begin{eqnarray}
	m_{\rm eff}^2 = \dfrac{2 N}{\beta} \rho_0^2 (\Lambda) + 2 (D-1) \left\langle R^2\right\rangle\,.
\end{eqnarray}
For the case when there is no symmetry breaking, it is easy to identify the usual extent of space parameter as 
\begin{equation}
\left\langle (R^i)^2\right\rangle := \frac{1}{N} \Tr \, \left\langle \left(X^i(\tau)\right)^2\right\rangle = \frac{N}{\beta} \sum_l\, v_{l} \, , 
\end{equation}
and is the same in all the $i$-directions\footnote{The case for cosmology is a bit more subtle. In that case, we find a symmetry breaking to give us three large spatial dimensions which are expanding; however, one does assume spatial isotropy amongst these large external directions. In essence there is an unbroken $SO(3)$ symmetry in that case when defining the extent of space parameters.}.

Before ending this section, let us note that the gap equation for the Gaussian parameters corresponding to the bosonic fields in the bosonic BFSS model is very similar to the one we had derived in the bosonic IKKT model. The appearance of the (infinite number of) Fourier modes does not actually complicate the story for the $SO(D)$ symmetry-breaking, \textit{i.e.} the $v_l$'s for all the different $i$'s are the same in the bosonic model. Thus, the main finding of this section is that the bosonic BFSS model has an unbroken $SO(D)$ symmetry, just like in the IKKT case. In hindsight, we could have guessed this conclusion from the results of \cite{Gaussian_methods_BFSS}; however, our calculation by not assuming an $SO(D)$-symmetric Gaussian action will be extremely helpful in the next section when including the fermionic terms.

\section{Symmetry breaking in the BFSS model}

The full BFSS model has the following additional terms due to the presence of fermionic matrices:
\begin{eqnarray}\label{BFSS_fermionic}
	S_{\rm BFSS}^{(f)} = \frac{i}{2 g^2} \sum_r \left(\frac{2\pi r}{\beta}\right) \, \psi_{-r} \psi_r - \frac{i}{2 g^2 \sqrt{\beta}} \sum_r \Tr \left(\psi_{-r} \left[A_{00}, \psi_r\right]\right)  - \frac{1}{2 g^2 \beta} \sum_{r,s} \Tr \left(\psi_{r} \Gamma_i \left[X_{-r-s}^i, \psi_s\right]\right)\,, 
\end{eqnarray}
where we have now Fourier expanded the fermionic fields as 
\begin{eqnarray}
	\psi_\alpha (\tau) = \sum_r \psi_r^\alpha e^{i r \omega \tau}\,.
\end{eqnarray}
We have suppressed the spinor index $\alpha$ throughout in \eqref{BFSS_fermionic} above. The Gamma matrices are $p \times p$ symmetric matrices satisfying the anti-commutation relations $\left\{\Gamma_i, \Gamma_j\right\} = 2 \delta_{ij}$. The above Fourier expanded terms comes from the two following fermionic terms in the Euclidean BFSS action:
\begin{eqnarray}\label{BFSS_fermionic1}
	S_{\rm BFSS}^{(f)} \propto \frac{1}{2 g^2} \int \d \tau \Tr \left( 
	\psi_\alpha D_\tau \psi_\alpha - \psi_\alpha \left(\Gamma_i\right)_{\alpha\beta}\, \left[X_i, \psi_\beta\right]\right)\,.
\end{eqnarray}

Including the fermionic terms in the BFSS action \eqref{BFSS_fermionic} requires adding to the Gaussian ansatz the following term:
\begin{eqnarray}\label{Gaussian_fermions}
	S_0^{(f)} = \sum_{\alpha, \beta}^p \sum_r \Tr \left(\psi_r^\alpha \mathcal{A}_{\alpha\beta} \psi^\beta_{-r}\right)\,, 
\end{eqnarray}
where we use the same symbol for the Gaussian parameter matrix $\mathcal{A}$ as we had done for the IKKT model. Recalling the wisdom gained from the IKKT case, we do not try to expand this matrix in terms of Gaussian parameters $\omega_{\mu\nu\lambda}$ and $\omega_\mu$ since this was not important to explore the question of existence of symmetry-breaking in the model, and unnecessarily complicates the gap equations. 

Before going through with the explicit calculations for the fermionic terms, let us make the following initial observations:
\begin{itemize}
 \item As we see from \eqref{BFSS_fermionic}, the fermions do not interact with the ghost fields and therefore, the gap equation for the ghost propagator will obviously remain decoupled. 
 \item On the other hand, we find that the fermions couple to the bosonic fields $X_l^i, A_{00}$ through cubic interactions. However,  and this the most subtle part of the calculation, although cubic terms of the form $\Tr \left(X_l^i\left[A_{00}, X_{-l}^i\right]\right)$ and $\Tr \left(\bar{\alpha_l}\left[A_{00}, \alpha_l\right]\right)$ do not contribute to the free energy (to the order we are interested in), as shown in the previous section, things are a bit different with fermionic terms. In fact, $\left\langle\left(S_{\rm BFSS}^{(f)}\right)^2\right\rangle$ does contribute to $F_1$ \cite{GE_fermion}. Thus, the cubic interactions, involving the fermion bilinears, will end up affecting the gap equation for the bosonic Gaussian parameters ($v_{l,i}$).
 \item The purely fermionic terms, on the other hand, can \textit{only} contribute to the gap equation for the fermionic parameters alone. So, we will ignore the contribution and focus only in the interaction terms which will have an effect on the $v_{l,i}$-gap equations.
\end{itemize}

Keeping the above general comments in mind, we will not try to calculate the full expression for the free energy, up to the next-to-leading order, including the fermionic terms. Instead, we will only try to identify the contribution which effect the $v_{l,i}$-gap equation \eqref{BFSS_bos_gap1}. The main ingredients required for doing this calculation are the fermionic propagators, calculated with respect to the Gaussian action \eqref{Gaussian_fermions}:
\begin{eqnarray}\label{Gaussian_prop_f}
	\left\langle\Big(\psi_r^\alpha\Big)_{AB} \left(\psi_s^\beta\right)_{CD}\right\rangle_0 = \left(\mathcal{A}^{-1}\right)^{\alpha\beta}\, \delta_{r,-s}\,\delta_{AD}\,\delta_{BC}\,.
\end{eqnarray}
As before, we will never need the expression for the propagator for the fermions calculated using the BFSS action itself, to the order we are considering, but we present that result here for completeness:
\begin{eqnarray}
	\left\langle\Big(\psi_r^\alpha\Big)_{AB} \left(\psi_s^\beta\right)_{CD}\right\rangle = \frac{g^2}{2\pi i r}\, \delta^{\alpha\beta}\, \delta_{r,-s}\,\delta_{AD}\,\delta_{BC}\,.
\end{eqnarray}

We begin with the following term: $-\frac{i}{2 g^2 \sqrt{\beta}} \, \sum_r \Tr\left(\psi_{-r} \left[A_{00}, \psi_r\right]\right)$, whose contribution to the free energy will scale as $\left\langle\left(S_{\rm BFSS}^{(f)}\right)^2\right\rangle$. Using the propagators from above, we can evaluate the contribution of this term as 
\begin{eqnarray}
	&\sim& \sum_s \sum_r \, \Big(A_{00}\Big)_{AB} \, \Big(\psi_{r}^\alpha\Big)_{BC} \, \Big(\psi_{-r}^\beta\Big)_{CA} \, \delta_{\alpha\beta}\, \Big(A_{00}\Big)_{DE} \, \Big(\psi_{s}^{\alpha'}\Big)_{EF} \, \Big(\psi_{-s}^{\beta'}\Big)_{FD} \, \delta_{\alpha' \beta'}\nonumber\\
	&\sim&  \rho^2\, \delta_{AE}\, \delta_{BD} \, \delta_{DB} \, \delta_{CF} \, \delta_{FC} \, \delta_{EA}\, \left[\left(\mathcal{A}\right)^{-1}_{\alpha\beta}\, \delta^{\alpha\beta}\right]^2\nonumber\\
	&\sim& \rho^2 \, N^3 \, \left[\Tr \left(\mathcal{A}^{-1}\right)\right]^2\,.
\end{eqnarray}
Although we have dropped a lot of numerical prefactors in the above calculation, the important conclusions are the following:
\begin{enumerate}
	\item There are no free $SO(D)$ or $(i)$ indices in the above expression, as is expected from the structure of the term itself. This clearly implies that even if the above term is to somehow influence the gap equation for $v_{l,i}$, it will surely allow for a $SO(D)$-symmetric solution of the form $v_{l,1} = v_{l,2} = \cdots = v_{l,D}\,,\; \forall \;l \in \mathbb{Z}$.
	\item However, more specifically in this case, the above term does not contain any factors of $v_{l,i}$, and can therefore not appear in the gap equation for $v_{l,i}$. The only way it can influence this gap equation is through $\rho_0$. Since the above equation does depend on $\rho_0$, and the latter appears in the $v_{l,i}$-gap equation, it can indirectly affect the solutions of $v_{l,i}$ through $\rho_0$. However, due to the argument mentioned above, it will not have any effect on the symmetry-breaking pattern.
	\item This small calculation demonstrates that \textit{any} type of a fermionic term will not lead to a $SO(D)$ symmetry breaking. In other words, having a matrix model action involving fermionic terms does not guarantee any symmetry-breaking and the structure of the Lagrangian itself is very important in exploring the pattern of symmetry-breaking.
\end{enumerate}

We are now finally in the position to tackle the main term that will give us evidence of symmetry-breaking in the BFSS model, namely the term
\begin{equation}
-\frac{1}{2 g^2 \beta}\,\sum_{r,s} \, \Tr \left(\psi_r \Gamma_i \left[X^i_{-r-s}, \psi_s\right]\right) \, .
\end{equation}
 Notice the striking similarity of this term with the (only) fermionic term in the IKKT model, which was ultimately responsible for symmetry-breaking in that case. Remembering that its contribution to the free energy is going to be at the quadratic order, we find
\begin{eqnarray}
	& & \sum_{i=1}^{D} \sum_{j=1}^{D} \sum_{p,q,r,s} \left\langle \Big(\psi_r^\alpha\Big)_{AB} \, \left(\Gamma_i\right)_{\alpha\beta}\, \left(X_{-r-s}^i\right)_{BC}\, \Big(\psi_s^\beta\Big)_{CA} \, \Big(\psi_p^{\alpha'}\Big)_{DE}\,  \left(\Gamma_j\right)_{\alpha\beta}\, \left(X_{-p-q}^j\right)_{EF}\, \Big(\psi_r^\alpha\Big)_{FD} \right\rangle\nonumber\\
	&\sim& \sum_{i=1}^{D} \sum_{j=1}^{D} \sum_{p,q,r,s}\, \delta_{r,-q} \,  \delta_{s,-p} \,  \delta_{-r-s+p+q} \,  \delta^{ij}\, N^3 \, v_{(-r-s), i}\, \left(\mathcal{A}^{-1}\right)^{\alpha\beta'}\, \left(\Gamma_i\right)_{\alpha\beta}\, \left(\mathcal{A}^{-1}\right)^{\beta\alpha'}\, \left(\Gamma_i\right)_{\alpha'\beta'}\nonumber\\
	&\sim& \sum_{i=1}^{D} \sum_{p,q} \, N^3\, v_{(p+q), i} \, \Tr \left[\left(\mathcal{A}^{-1}\, \Gamma_i\right)^2\right]\,.
\end{eqnarray}
Since $(p, q) \in \mathbb{Z}/2$, and we have an infinite sum over both, we can replace this index by some $l \in \mathbb{Z}$, such that we now have
\begin{eqnarray}
	\sum_{i=1}^{D} \sum_{l} \, N^3\, v_{l, i} \, \Tr \left[\left(\mathcal{A}^{-1}\, \Gamma_i\right)^2\right]\,.
\end{eqnarray}
This, of course, depends explicitly on $v_{l,i}$ and would therefore contribute to the gap equation of interest. In fact,  its contribution to the gap equation will be a term of the form 
\begin{equation}
\frac{N}{g^2\beta}\, \Tr \left[\left(\mathcal{A}^{-1}\, \Gamma_i\right)^2\right] \, . 
\end{equation}
When we include this in \eqref{BFSS_bos_gap1}, we get an equation of the form:
\begin{eqnarray}\label{Main_result}
		- \frac{1}{v_{l,i}} + \frac{1}{g^2} \left(\dfrac{2\pi l}{\beta}\right)^2 + \dfrac{2 N}{g^2 \beta}\, \rho_0^2 (\Lambda)  + \dfrac{2 N}{g^2 \beta} \,\sum_{\j\neq i} \sum_l \, v_{l,j} - \frac{N}{g^2\beta}\, \Tr \left[\left(\mathcal{A}^{-1}\, \Gamma_i\right)^2\right] = 0\,.
\end{eqnarray}
The last term shows that this equation cannot have an $SO(D)$ symmetric solution as long as the matrix $\mathcal{A}$ has at least one non-zero entry. This is analogous to saying that the fermionic Gaussian parameters are not trivially zero, and for the reasons emphasized earlier for the IKKT model, we are assured that the solution to the above equation must break $SO(D)$ symmetry.

\section{Conclusion}

The recent developments in matrix cosmology have provided a promising new direction in understanding our early universe. For the IKKT model, it has been shown \cite{us2} how analytical methods can be used to extract a spacetime metric, with an infinite extent for both space and time, from the (block-diagonal) structure and dynamics of the matrices. This was shown to naturally solve the flatness problem of standard big bang cosmology, and hint towards a solution for the cosmological constant issue. On the other hand, a thermal state in the BFSS model has been shown to provide a natural solution to the horizon problem as well as predict an almost scale-invariant spectrum of primordial perturbations \cite{us1}.  A crucial input input the scenario of \cite{us1} was the assumption that the spatial rotational symmetry of the BFSS Lagrangian is spontaneously broken in the state which minimizes the free energy. to a configuration in which the extent of space only becomes large in three spatial directions.  The existence of such a phase transition has been established in the IKKT model, but not yet in the BFSS scenario.

In this paper, we have provided first  evidence for a $SO(9)$ symmetry-breaking in the BFSS model, similar to what happens in the IKKT case, by employing the Gaussian expansion method (the same methods which were used in the case of the IKKT model to show the existence of the symmetry breaking phase transition).  The inclusion of the contribution of fermions is crucial to reach this conclusion.  In the absence of fermions, the state which minimizes the free energy maintains the $SO(9)$ symmetry.   In the case of the IKKT model, numerical studies (both full matrix model simulations and also numerical evaluations of the free energies) have shown that the energetically favored state has $SO(3)$ symmetry with three dimensions of space becoming large.  A next step in our research program is to perform similar analyses in the case of the BSFF model in order to determine the symmetry and features of the configuration after the breaking of the $SO(9)$ symmetry.  This would require going well-beyond the next-to-leading order calculations done here and cannot be achieved by analytical tools alone. However, since the corresponding calculations for the IKKT model have been manageable, it is only natural to push for examining if such a similar result can be obtained for the BFSS model. If possible, such a result would be  prove that a large $(3+1)$-d universe can spontaneously emerge in the BFSS matrix model. 

There has already been a wealth of similarities between results coming out of matrix cosmology and string gas cosmology. For instance, the amplitude of (scale-invariant) perturbations for the thermal state in the BFSS model is exactly the same as in string gas cosmology (see e.g. \cite{SGCrev} for a review).  It is well known that three large dimensions do emerge in the string gas model since space cannot expand unless the winding modes of the strings annihilate into string loops, the probability for which is zero only if there are more than three large spatial dimensions. We believe that this striking similarity in explaining the emergence of a large $(3+1)$-d universe from full string theory, in both string gas cosmology and matrix models, is not a mere coincidence and that the physical reason underlying both must be the same. In fact, there must be a well-defined sense in which one can recover the string gas model from the full dynamics of matrix theory. Another goal for the future will be to further explore the physical reason behind the SSB in the BFSS and IKKT models since this might point to the aforementioned connection.

\bigskip

\section*{Acknowledgements:}

SB is supported in part by the Higgs Fellowship. SL is supported in part by FRQNT. The research at McGill is supported in part by funds from NSERC and from the Canada Research Chair program.  \\
For the purpose of open access, the authors have applied a Creative Commons Attribution (CC BY) licence to any Author Accepted Manuscript version arising from this submission.

\bigskip

\appendix

\section{Computation of the free energy for the bosonic BFSS action} 
In the appendix, we give details of the computation of the free energy, up to the first order, for the bosonic BFSS model which has been used in the main draft to derive the gap equation for the bosonic Gaussian parameters.

\subsection{Derivation of $F_0$}\label{App1}

We want to find the expressions for all the terms appearing in \eqref{F0_bosonic}, which we reproduce here for convenience:
\begin{eqnarray}
		\beta F_0 = \beta F_\square(\Lambda) - \frac{N^2}{2}\sum_{i=1}^{D} \sum_l \ln v_{l,i} + N^2 \sum_{l\neq 0} \ln s_l\,.
\end{eqnarray}
The first term is what it is by definition, and its explicit form can be read from \eqref{Hol_1}. Let us begin by deriving the second term on the r.h.s. of the equation above which involves the second term in the Gaussian ansatz \eqref{BFSS_Gaussian_bosonic}:
\begin{eqnarray}
S_0^{\rm bosonic} &=&\sum_l \sum_{i=1}^{D} \frac{1}{2 v_{l,i}} \, \Tr \left(X^i_l X^i_{-l}\right) \nonumber\\
 &=&\sum_l \sum_{i=1}^{D} \sum_{a=1}^{N^2}\frac{1}{2 v_{l,i}} \, \Tr \left(X^{i\,a}_l X^i_{-l\, a}\right)\,,
\end{eqnarray}
where we have explicitly written the trace in terms of the $U(N)$ generators, and $a$ refers to the $U(N)$ index. Given this action, the corresponding partition function (for the above term in the Gaussian action) is given by:
\begin{eqnarray}
	Z_0^{\rm bosonic} &=& \int \prod_{a=1}^{N^2} \, \prod_{i=1}^D\, \d x_i^a\, e^{-S_0^{\rm bosonic}}\\
	&=&\int \prod_{a=1}^{N^2} \, e^{-\sum_l \frac{1}{2 v_{l,1}}\left(x_1^a x_{1\,a}\right)} \; \int \prod_{a=1}^{N^2} \, e^{-\sum_l \frac{1}{2 v_{l,2}}\left(x_2^a x_{2\,a}\right)}\,\ldots\, \int \prod_{a=1}^{N^2} \, e^{-\sum_l \frac{1}{2 v_{l,D}}\left(x_D^a x_{D\,a}\right)}\nonumber\\
	&=&  \underbrace{\left(2 \sum_l v_{l,1}\right)^{N^2/2}\;\left(2 \sum_l v_{l,2}\right)^{N^2/2}\,\ldots\, \left(2 \sum_l v_{l,D}\right)^{N^2/2}}_{D\;{\rm number\; of\; terms}}\,,\nonumber
\end{eqnarray}
where we have omitted several factors of numerical constants (involving $\pi$ in the multiple Gaussian integrals) as this will only give a constant contribution to the free energy which is irrelevant for the gap equation. We have also suppressed a factor of $\beta$ which we will restore later by noting that $Z \sim e^{-\beta S}$. Thus, a factor of $\beta$ will appear in the denominator after carrying out the Gaussian integrals and will finally cancel with the $\beta$ from $\beta F_0$. 

Moving forward, the free energy corresponding to this is given by
\begin{eqnarray}
	F_0^{\rm bosonic} &=& -\ln Z_0^{\rm bosonic}\\
	&=&   -\frac{N^2}{2} \left[\sum_l \ln v_{l,1} + \sum_l \ln v_{l,2} + \ldots + \sum_l \ln v_{l,d}\right] = -\frac{N^2}{2} \sum_{i=1}^{D} \sum_l \ln v_{l,i}\,.
\end{eqnarray}
A similar analysis for the ghost fields yields:
\begin{eqnarray}
	F_0^{\rm ghost} = N^2 \sum_{l\neq 0} \ln s_l\,. 
\end{eqnarray}

\subsection{Derivation of $F_1$}\label{App2}

Let us calculate this term by term for the (bosonic part of the) BFSS action given in \eqref{BFSS_bosonic}, as well as the Gaussian action, which we rewrite below and label the different terms, as follows:
\begin{eqnarray}\label{BFSS_bosonic2}
	S_{\rm BFSS}^{(b)} &=& \underbrace{\frac{1}{2 g^2} \sum_l \left(\dfrac{2\pi l}{\beta}\right)^2 \Tr\left(X_l^i X_{-l}^i\right)}_{(1)} \;+\; \underbrace{\frac{1}{g^2} \sum_{l\neq 0} \left(\dfrac{2\pi l}{\beta}\right) \Tr\left(\bar{\alpha}_l \alpha_{l}\right)}_{(2)}\\
	& & - \underbrace{\frac{1}{g^2 \sqrt{\beta}}\sum_l \left(\dfrac{2\pi l}{\beta}\right)^2 \Tr\left(X_l^i \left[A_{00}, X_{-l}^i\right]\right) + \frac{1}{g^2 \sqrt{\beta}}\sum_{l\neq 0} \left(\dfrac{2\pi l}{\beta}\right) \Tr\left(\bar{\alpha}_l \left[A_{00}, \alpha_{-l}\right]\right)}_{(6)} \nonumber\\
	& & +  \underbrace{\frac{1}{2 g^2 \beta}\sum_l \Tr\left(\left[A_{00}, X_l^i\right] \left[A_{00}, X_{-l}^i\right]\right)}_{(3)} \;-\;   \underbrace{\frac{1}{4 g^2 \beta}\sum_{l+m+n+p=0} \Tr\left(\left[X_l^i, X_m^j\right] \left[X_n^i, X_p^j\right]\right)}_{(4)}\nonumber\,.	
\end{eqnarray}

\begin{eqnarray}\label{BFSS_Gaussian_bosonic1}
	S_0^{(b)} = - \underbrace{\frac{N}{\Lambda} \Tr \left(U + U^\dagger\right)}_{(5)} \;+\; \underbrace{\sum_l \sum_{i=1}^{D} \dfrac{1}{2 v_{l,i}} \Tr \left(X_l^i X_{-l}^i\right)}_{(1)} \; -\;  \underbrace{\sum_{l\neq 0}\frac{1}{s_l} \Tr\left(\bar{\alpha}_l\alpha_l\right)}_{(2)}\,,
\end{eqnarray}
We will repeatedly use the propagators written down in \eqref{Gaussian_prop_bosonic} in order to do the explicit calculation. As mentioned earlier, and shown below, we will never need to consider the propagators given in \eqref{BFSS_bosonic_prop} for evaluating $F_1$.

The easiest to calculate are the $(5)$ terms involving the Wilson loop operators which are given by
\begin{eqnarray}
	 \left\langle \left(S-S_0\right)\right\rangle_0 = \frac{N}{\Lambda} \left\langle \Tr\, \left(U + U^\dagger\right)\right\rangle_\square\,,
\end{eqnarray}
the expression for which has been given in \eqref{F1_hol}. 

Next we focus on the terms labelled by $(1)$:
\begin{eqnarray}
	\left\langle S \right\rangle_0 \sim \sum_{i=1}^D \sum_l \frac{1}{2 g^2} \left(\frac{2\pi l}{\beta}\right)^2\, v_{l,i} \, \delta_{AA} \delta_{BB} = \sum_{i=1}^D \sum_l \frac{N^2}{2 g^2} \left(\frac{2\pi l}{\beta}\right)^2\, v_{l,i}\,,
\end{eqnarray}
and
\begin{eqnarray}
	\left\langle S_0 \right\rangle_0 \sim  \sum_{i=1}^D \sum_l  \frac{1}{2 v_{l,i}} \left\langle  \Tr \left(X_l^i X_{-l}^i\right) \right\rangle_0 = \dfrac{N^2}{2} \sum_{i=1}^D \sum_l  \,1\,.
\end{eqnarray}
Although we could have carried out the sum over $i$ in the second term above, we keep it in this form since in this way it is easier to organize the terms later on.

The terms labelled by $(2)$ can be evaluated as:
\begin{eqnarray}
\left\langle S \right\rangle_0 \sim \frac{N^2}{g^2} \sum_{l \neq 0} \left(\frac{2\pi l}{\beta}\right)^2 \, s_l \,,
\end{eqnarray}
and 
\begin{eqnarray}
	\left\langle S_0 \right\rangle_0 \sim N^2 \sum_{l \neq 0} \,1 \,.
\end{eqnarray}
Note that naively the sum over the Fourier modes give an infinite contribution to each of these terms, but this is not a problem for us since the terms are independent of the bosonic gap parameters.

Although the terms marked $(3)$ look more complicated, they can easily be evaluated by keeping in mind the following considerations. Firstly, there are no such quartic terms in the Gaussian ansatz \eqref{BFSS_Gaussian_bosonic1}. And secondly, we need to only calculate the dominant (connected) term in the large-N limit (which corresponds to choosing the right contractions of the $U(N)$ indices $A,B,C,D$ etc.):
\begin{eqnarray}
	\left\langle S \right\rangle_0 \sim \frac{1}{g^2 \beta} \sum_{i=1}^D \sum_l \, 	\left\langle (A_{00})_{AB} (A_{00})_{BC} (X_l^i)_{CD} (X_{-l}^i)_{DA} \right\rangle_0 = \frac{N^3}{g^2 \beta} \,\rho_0^2 \left(\Lambda \right) \, \sum_{i=1}^D \sum_l v_{l,i}\,.
\end{eqnarray}

Similarly, the term marked $(4)$ can be evaluated to be:
\begin{eqnarray}
	\dfrac{N^3}{2 g^2 \beta}\, \underbrace{\sum_{i=1}^D \sum_{j=1}^D}_{j\neq i} \sum_l\, v_{l,i}\, v_{l,j}\,.
\end{eqnarray}

Finally, the terms marked $(6)$ actually do not give any contribution to the free energy to the order we are considering at all. This comes from the simple observation that these terms are cubic and thus have zero contribution for the (quadratic) Gaussian propagators. However, the important observation in this regard is that this conclusion is only true for the $X_l^i, A_{00}$ and $\alpha_l$ fields since they are all bosonic in nature, and would not be applicable for the fermionic fields as we will see later on.

\end{document}